\documentclass[10pt,letterpaper]{article}
\usepackage[top=0.85in,left=2.75in,footskip=0.75in]{geometry}
\usepackage{algorithm}
\usepackage{algpseudocode}
\usepackage[dvipsnames]{xcolor}

\usepackage{amsmath,amssymb}
\usepackage{graphicx}
\usepackage{subcaption}
\usepackage{caption}

\usepackage[percent]{overpic}

\usepackage{changepage}
\usepackage{float}

\usepackage{textcomp,marvosym}

\usepackage{cite}
\usepackage{booktabs}
\usepackage{nameref,hyperref}

\usepackage[expansion=false]{microtype}

\DisableLigatures[f]{encoding = *, family = * }

\usepackage[table]{xcolor}

\usepackage{array}

\newcolumntype{+}{!{\vrule width 2pt}}

\newlength\savedwidth

\usepackage[aboveskip=1pt,labelfont=bf,labelsep=period,justification=justified,singlelinecheck=off]{caption}

\makeatletter
\renewcommand{\@biblabel}[1]{\quad#1.}
\makeatother

\usepackage{lastpage,fancyhdr,graphicx}
\usepackage{epstopdf}

\fancyheadoffset[L]{2.25in}
\fancyfootoffset[L]{2.25in}
\makeatletter
\def\showhyphens#1{}%
\makeatother

\usepackage{algorithmicx}

\usepackage[most]{tcolorbox}

\algnewcommand{\Input}{\State \textbf{Input:} }
\algnewcommand{\Output}{\State \textbf{Output:} }
\algnewcommand{\Initialize}{\State \textbf{Initialize:} }

\begin{document}
\vspace*{0.2in}

% Title must be 250 characters or less.
\begin{flushleft}
{\LARGE
\textbf\newline{Social norm dynamics in a behavioral epidemic model} 
}
\newline

Christos Charalambous\textsuperscript{1*}

\bigskip
\textbf{1} University of Cyprus, Department of Economics, PO Box 20537, 1678 Nicosia, Cyprus
\\
\bigskip

* charalambous.christos.2@ucy.ac.cy

\end{flushleft}
% Please keep the abstract below 300 words
\section*{Abstract}

Understanding the social determinants of preventive behavior is vital for epidemic modelling and effective policy making. Traditional models emphasize imitation or rational trade-offs, but recent evidence highlights the role of social norms. We develop a behavioral epidemic model of seasonal disease on multilayer networks, where vaccination decisions combine learning from experience with coevolving social norms. The framework distinguishes descriptive norms (what others do) from injunctive norms (what others think ought to be done), while incorporating cognitive dissonance, social projection and logical consistency. Simulations show that norm dynamics yield markedly different vaccination uptake and infection levels compared to considering solely payoff-driven learning. Injunctive norms exert stronger and more persistent effects than descriptive norms. Interventions targeting injunctive expectations improve outcomes, while those on descriptive norms may be weaker or even counterproductive. Norm-based models, once empirically validated, can better capture human behavior and guide strategies for collective action problems even beyond pandemics.

\flushbottom

\thispagestyle{empty}

\section*{Introduction}

Vaccine hesitancy is one of the most severe major global health threats \cite{2019WHO}, made more pressing by the experience of COVID-19 \cite{2023Vriens} and the persistence of seasonal influenza. As with many collective-action problems, relying solely on regulation is often too slow or too costly \cite{2009Walker}. This has motivated a growing interest in behavioural approaches that draw on social norms to encourage cooperation and curb free-riding \cite{2016Nyborg}.

Models that couple disease dynamics with human behaviour describe how infection and protective actions influence one another \cite{2015Wang,2013Granell,2024Han}, while game-theoretic approaches account for strategic vaccination decisions made under bounded rationality \cite{2004Bauch,2010Reluga,2019ShiA,2011Bhattacharyya}. However, these models usually capture social influence in overly narrow ways—most often as simple imitation—thereby missing important psychological mechanisms such as cognitive dissonance \cite{1957Festinger}, social projection \cite{2007Krueger}, and higher-order expectations \cite{2010Apperly}, all of which shape how people form and revise their beliefs and intentions \cite{1990Cialdini,2005Bicchieri}. To reflect real decision-making, models need to distinguish between descriptive norms—what others do—and injunctive norms—what others think should be done \cite{1990Cialdini}. When these two coincide, compliance and cooperation are typically much stronger \cite{2005Bicchieri,2000Ostrom}.

Empirical evidence consistently shows that pro-vaccination social norms lead to higher vaccine uptake across different diseases \cite{2021Agranov,2021Graupensperger,2023Moehring}, and public-health strategies increasingly focus on interventions that harness these norms \cite{2017Brewer}. In most modelling approaches, descriptive norms are introduced through social learning processes \cite{2012Bauch,2012dOnofrio}, while injunctive norms—anchored in moral and reputational expectations—have received much less attention \cite{2014Oraby,2022Battigalli,2008LopezPerez}, even though they are crucial for sustaining long-term behavioural change. Only a few models have attempted to integrate the two \cite{2018RealpeGomez1}, and more recent psychologically grounded formulations now offer a firmer foundation for doing so \cite{2021Gavrilets}.

Most behavioural–epidemic models represent social influence through imitation or by averaging payoffs, treating norms as fixed factors that modify utility rather than as evolving psychological states \cite{2004Bauch,2010Reluga,2019ShiA}. In contrast, the framework introduced here distinguishes between personal norms, empirical expectations, and normative expectations, linking their joint evolution to experience-weighted learning. This approach moves beyond imitation and payoff-based mechanisms, allowing us to explore how multiple layers of social norms coevolve with epidemic risk. 
%As a result, the model can reproduce dynamic patterns—such as sustained vaccination even at low infection levels and varied responses to external interventions—that standard formulations fail to capture.

An expanding line of research is now developing cognitively grounded, empirically informed agent-based models (ABMs) to investigate how norms emerge, exert downward causation, and shape long-term intervention outcomes \cite{2021Li,2012Bravo,2021Taghikhah,2018RealpeGomez,1974Campbell,1998Castelfranchi}. In their “transmission game,” Woike et al. \cite{2022Woike} found that communicating injunctive norms significantly reduced risk-taking, while merely providing descriptive information or case counts had little effect—or even backfired. Building on this evidence, we introduce an ABM that integrates a behavioral epidemic model with realistic social-norm dynamics \cite{2021Gavrilets}, following the approach of Realpe-Gómez et al. \cite{2018RealpeGomez}, to explore how vaccination intentions and norms evolve over successive seasonal outbreaks.

We first describe the model—its decision process, Experience-Weighted Attraction (EWA) learning \cite{1999Camerer}, and norm dynamics—and then analyse how social norms influence infection and vaccination patterns, including cases of stubborn adherence and external interventions. The framework is suitable for future laboratory validation \cite{2012Chapman}.

Our contributions are fourfold: (i) we couple epidemic and norm dynamics through adaptive learning weights \cite{2007Ho,2023Heiman,2015Gelfand}; (ii) we integrate cognitive dissonance, social projection, and logical consistency into norm evolution \cite{2014Oraby,2021Gavrilets}; (iii) we show that injunctive norms exert stronger long-term effects on vaccination than descriptive ones, consistent with experimental findings \cite{2021Szekely,2022Woike}; and (iv) we model external interventions acting directly on norms rather than merely on information flows \cite{2024Meiske,2022Tverskoi,2022Gavrilets}.
\vspace{-0.5em}
\section{Model}

During a pandemic, two coupled processes unfold: infection spreads through contact, and individuals adapt behaviour in response. The epidemic evolves according to a Susceptible–Infected–Recovered (SIR) process on a physical contact network, while behavioural change occurs on an overlapping social layer.

\begin{figure}[H]
  \centering

  % ---------- (a) Algorithm panel ----------
  \begin{subfigure}[t]{0.50\textwidth}  % <--- smaller than 0.5 to help them fit
    \vspace{0pt}
    \begin{tcolorbox}[
      colback=white,
      boxrule=0.3pt,
      left=6pt,right=6pt,top=6pt,bottom=6pt,
      enhanced,
      overlay={\node[anchor=north west, font=\bfseries] at (frame.north west) {a};}
    ]
      \begin{algorithmic}[1]
        \Input Physical network $G^{\text{phys}}$, social network $G^{\text{soc}}$, SIR parameters $(\beta,\mu)$, memory $m$
        \Output Seasonal infections $I(t)$, vaccination probabilities $x_i(t)$, and norms $\{y_i(t),\tilde{x}_i(t),\tilde{y}_i(t)\}$
        \Initialize epidemic and normative states.
        \For{$t = 0$ to $T_{\max}$}
          \State \textbf{Epidemic season:} run $n_{\text{sim}}$ SIR realizations on $G^{\text{phys}}$ with actions $a(t)$; record $\hat{I}_i(t)$ and $\hat{I}_{i,\text{neigh}}(t)$.
          \State \textbf{Safety weight:}
          \[
            S_i(t)=
            \begin{cases}
              1-\hat{I}_i(t), & a_i(t)=0\\
              1-\hat{I}_{i,\text{neigh}}(t), & a_i(t)=1
            \end{cases}
          \]
          \State \textbf{Peer summary \& trust:} compute $X_i(t)$; update $\phi_{\text{change},i}(t)$, $\phi_{\text{cons},i}(t)$, $\phi_i(t)=\sqrt{\phi_{\text{change},i}(t)\,\phi_{\text{cons},i}(t)}$.
          \State \textbf{Empirical channel:} $x^{\text{emp}}_i(t)=(1-\phi_i(t))\,p^{\text{learn}}_i(t)+\phi_i(t)\,X_i(t)$.
          \State \textbf{Injunctive channel:} $x^{\text{inj}}_i(t)=(1-\phi_i(t))\,y_i(t)+\phi_i(t)\,\tilde{y}_i(t)$.
          \State \textbf{Update:} $x_i(t+1)=(1-S_i(t))\,x^{\text{emp}}_i(t)+S_i(t)\,x^{\text{inj}}_i(t)$; draw $a_i(t+1)\sim \mathrm{Bernoulli}(x_i(t+1))$.
          \State \textbf{Norm updates (de-Groot)} \& \textbf{Convergence test}.
        \EndFor
      \end{algorithmic}
    \end{tcolorbox}
    \caption{}
    \label{fig:Fig1a}
  \end{subfigure}%
  \hfill
  % ---------- (b) Network + schematic ----------
  \begin{subfigure}[t]{0.48\textwidth}
    \vspace{5em}
    \centering

	\begin{tcolorbox}[
      colback=white,
      boxrule=0.3pt,
      left=6pt,right=6pt,top=6pt,bottom=6pt,
      enhanced,
      overlay={\node[anchor=north west, font=\bfseries] at (frame.north west) {};}
    ]    
    
    \begin{overpic}[width=\linewidth]{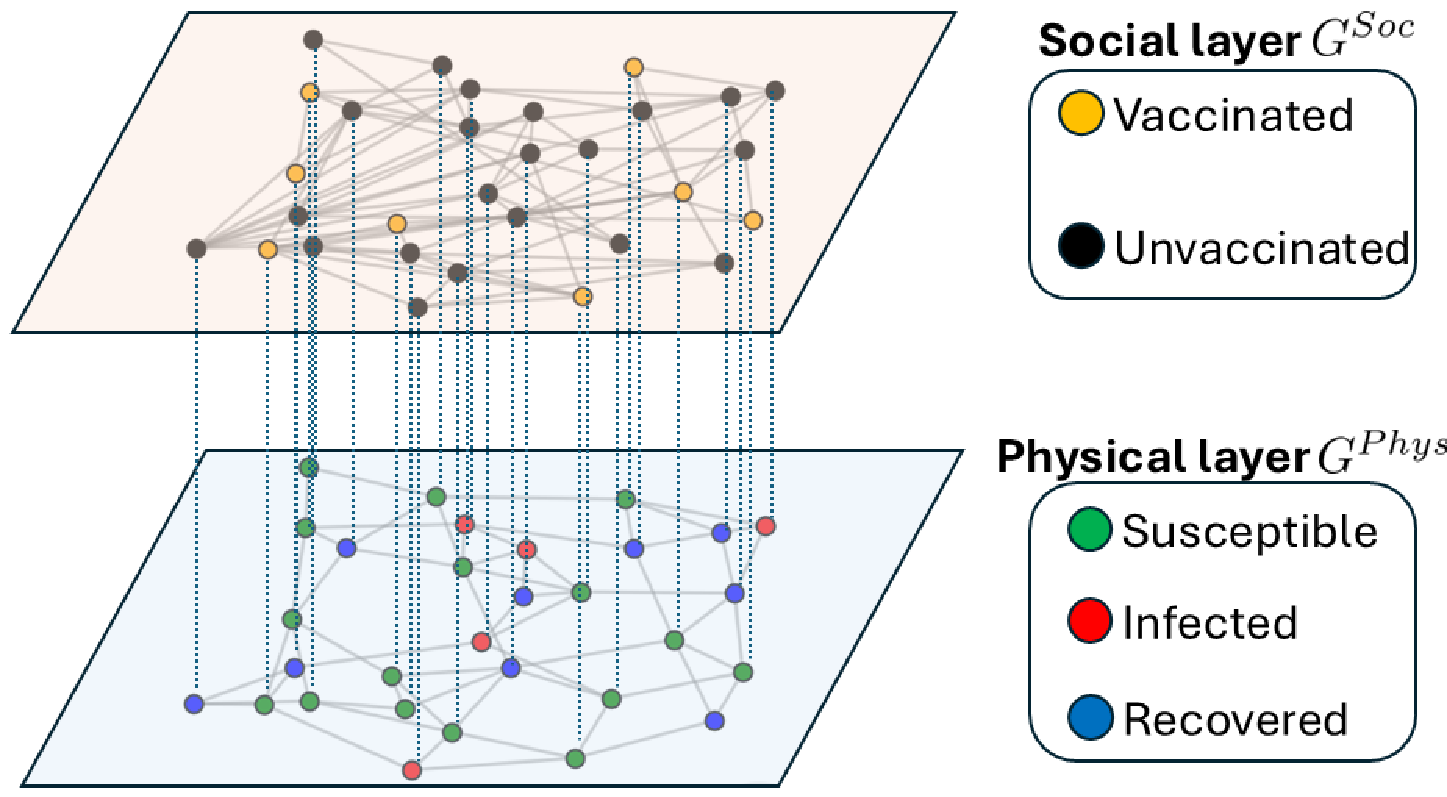}
    %\vspace{1.5pt}
      \put(0,55){\bfseries b} % top-left label
    \end{overpic}\\[4.5em]
    \includegraphics[width=\linewidth]{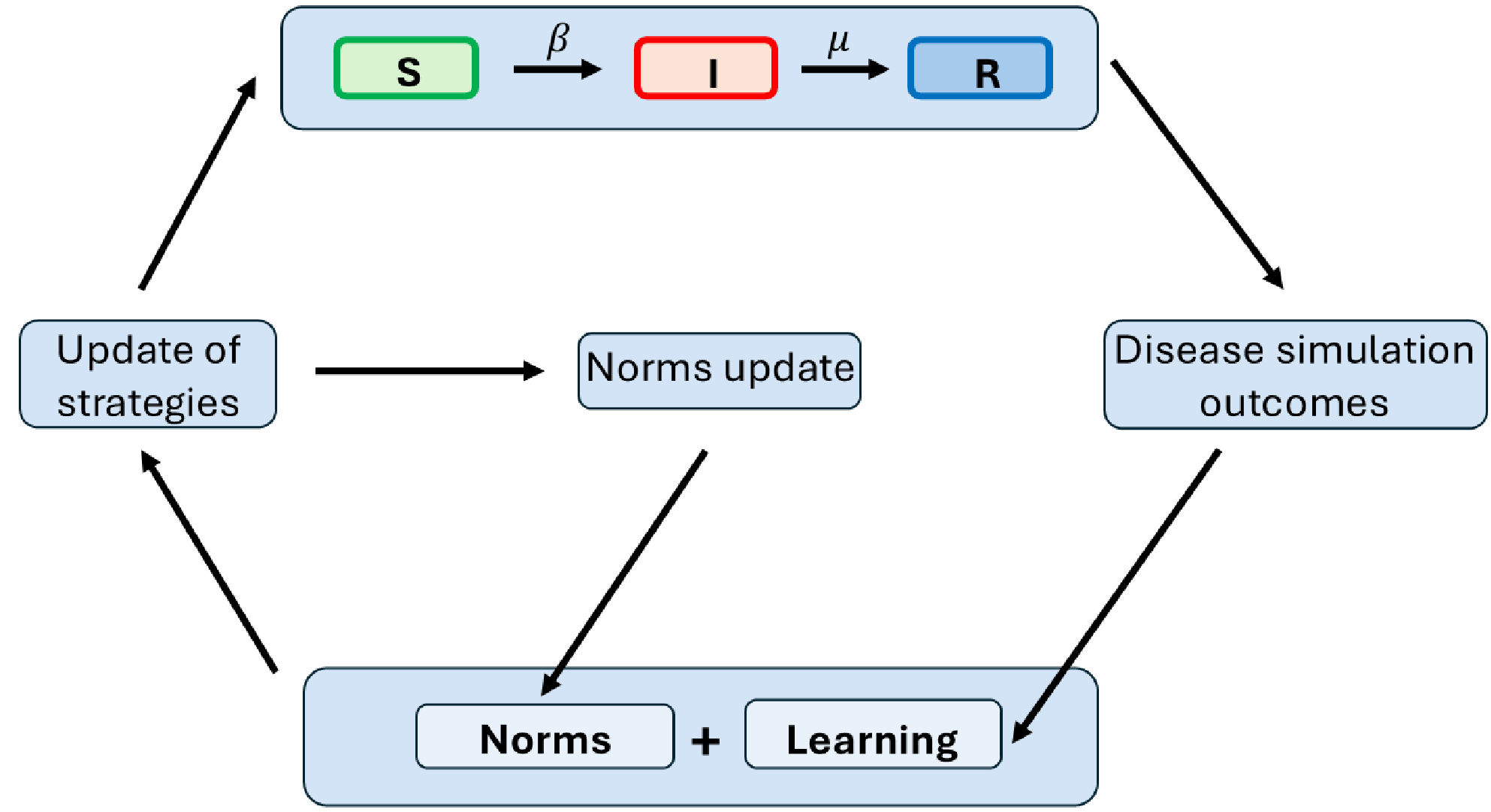}\\[2.5em]
    \caption{\textbf{Top:} Two-layer multiplex network on which the dynamics are implemented. 
             \textbf{Bottom:} Schematic of the dynamics.}
    \label{fig:Fig1b}

	\end{tcolorbox}    
    
  \end{subfigure}

  \caption{\textbf{Dynamics' algorithm, Two-layer network and schematic of dynamics. a.:} Algorithm illustrating the system dynamics. \textbf{b. Top:} Vaccination game on a two-layer network. The physical layer hosts a seasonal SIR epidemic, while the social layer governs vaccination decisions based on information from both layers. The physical layer follows a small-world topology and the social layer a Klimek–Thurner network, with partial overlap reflecting that many physical contacts also serve as information links. \textbf{Bottom:} Schematic of the algorithm. Each season, the SIR dynamics run on the physical layer to estimate infection probabilities. Agents then decide whether to vaccinate, considering both payoff-based learning and normative factors, update their norms accordingly, and begin the next season.}
  \label{fig:Fig1}
\end{figure}

We considered empirically motivated synthetic networks. The physical layer follows a small-world topology \cite{2010Salathe} with mean degree 6 and rewiring probability 0.1. The social layer follows a Klimek–Thurner network \cite{2013Klimek,2023Charalambous} (with  parameters as in the original work $c=0.58$, $r=0.12$ and $m=1$) and partial overlap with the physical layer, reflecting that many physical contacts also serve as information links. The results obtained were robust across other topologies as well, for example when both networks are characterized by an Erdos-Renyi topology. The two-layer network architecture is depicted in Fig.~\ref{fig:Fig1}.

Each season, agents decide whether to vaccinate \cite{2013Cardillo}; one agent starts infected, and the system evolves over successive seasons until equilibrium is reached. Each season comprises multiple SIR simulations using an event-driven algorithm \cite{2017Kiss} (with µ = 1) to estimate infection probabilities. Although social norms could in principle evolve continuously to reflect external shocks or information flow, our goal is to capture how individuals’ choices shape norm evolution. Therefore, norm updates occur only at the start of each season, when vaccination decisions are made. The algorithm and a comprehensive schematic of it are shown in Fig. \ref{fig:Fig1}.

\vspace{-0.5em}

Each season, agents decide whether to vaccinate \cite{2013Cardillo}; one agent starts infected, and the system evolves over successive seasons until equilibrium is reached. Each season comprises multiple SIR simulations using an event-driven algorithm \cite{2017Kiss} (with µ = 1) to estimate infection probabilities. Although social norms could in principle evolve continuously to reflect external shocks or information flow, our goal is to capture how individuals’ choices shape norm evolution. Therefore, norm updates occur only at the start of each season, when vaccination decisions are made. The algorithm and a comprehensive schematic of it are shown in Fig. \ref{fig:Fig1}.

To model behavioral adaptation, we employ a game-theoretic framework in which vaccination payoffs depend on others’ choices. Decisions are influenced not only by epidemiological risk but also by personal beliefs and social norms. Following \cite{2021Gavrilets}, each individual $i$ is characterized by (i) a personal attitude $y_i$ (belief about the appropriate behavior), (ii) an empirical expectation $\widetilde{x}_i$ (belief about others’ intentions), and (iii) a normative expectation $\widetilde{y}_i$ (belief about others’ attitudes).

\subsection*{The decision making process\label{subsec:The-decision-making}} We define the dynamics of an agent’s vaccination intention $x_i(t)$, i.e. the probability of choosing vaccination, which we assume to be driven by both material and normative considerations. Empirical evidence suggests that communities under risk rely more on experience and descriptive norms than on moral beliefs or injunctive norms \cite{2023Heiman,2015Gelfand}. In the absence of precise data on how perceived risk shapes this balance, we assume a simple linear relation between risk and the weight placed on empirical factors, following standard practice in norm–utility models \cite{2022Tverskoi}. Therefore we split the contribution to the intention $x_{i}\left(t+1\right)$
update as follows:

\begin{equation}
x_{i}\left(t+1\right)=\left(1-S_{i}\left(t\right)\right)x_{i}^{empirical}\left(t\right)+S_{i}\left(t\right)x_{i}^{injunctive}\left(t\right)
\end{equation}
which depends on the safety parameter $S_{i}\left(t\right)$ that
quantifies the safety an agent feels and takes values $0<S_{i}\left(t\right)<1$.
The safety is assumed to be given by the following formula

\begin{equation}
S_{i}\left(t\right)=\begin{cases}
1-\widehat{I}_{i}\left(t\right) & if\,agent\,unvaccinated\\
1-\widehat{I}_{i,neighbors}\left(t\right)& if\,agent\,vaccinated
\end{cases}
\end{equation}
where $\widehat{I}_{i}\left(t\right)=\frac{n_{sim}^{inf}\left(t\right)}{n_{sim}}$
with $n_{sim}$ the total number of outbreak simulations and $n_{sim}^{inf}\left(t\right)$
the number of outbreak simulations when $i$ got infected in the previous
cycle, and $\widehat{I}_{i,neighbors}\left(t\right)=\frac{n_{sim}^{inf,neigh}\left(t\right)}{n_{sim}}$
with $n_{sim}^{inf,neigh}\left(t\right)$ the sum of the average
number of infected neighbors over all simulation runs. 

We define the contribution from what is or what agents
think is actually happening, i.e. the empirical contribution to the intention, as

\begin{equation}
x_{i}^{empirical}\left(t\right)=\left(1-\phi_{i}\left(t\right)\right)p_{i}^{learn}\left(t\right)+\phi_{i}\left(t\right)\widetilde{x}_{i}\left(t\right)
\end{equation}
where we see contributions from the agent's learning of material payoffs
$p_{i}^{learn}\left(t\right)$ (see next subsection, $0\leq p_{i}^{learn}\left(t\right)\leq1$) and what the agent thinks that the
rest of the community actually chose on average $\widetilde{x}_{i}\left(t\right)$ (empirical expectations, $0\leq\widetilde{x}_{i}\left(t\right)\leq1$).
Their relative weight depends on $\phi_{i}\left(t\right)$, which determines whether decisions rely more on personal experience or on social information

\begin{equation}
\phi_{i}\left(t\right)=\sqrt{\phi^{change}_{i}\left(t\right)\phi^{consensus}_{i}\left(t\right)}.
\end{equation}
$\phi^{change}_{i}\left(t\right)$ is the change-detector function and is defined as in \cite{2007Ho}. $\phi^{consensus}_{i}\left(t\right)\in[0,1]$ is the consensus function given by
\begin{equation}
\phi^{consensus}_{i}\left(t\right)= |2*X_i-1|
\end{equation} 
where $X_{i}$ is the average
fraction of neighbors vaccinated as observed by the focal individual, and hence $\phi^{consensus}_{i}\left(t\right)$ is equal to 1 when neighbors agree and 0 if only half are vaccinated.

Intuitively, $\phi_{\mathrm{change},i}$ measures how volatile neighbours’ behaviour has been compared to the recent past, whereas $\phi^\mathrm{consensus}_{i}(t)$, captures the degree of agreement among them. Using the geometric mean ensures that social information becomes influential only when both conditions are met—behaviour is stable and neighbours largely agree—whereas in noisy or polarized environments agents fall back on their own experience and attitudes.

\begin{figure}[H]
  \centering

  % ---- Subfigure (a)
  \begin{subfigure}[t]{0.525\textwidth}
    \centering
    \begin{overpic}[width=\linewidth]{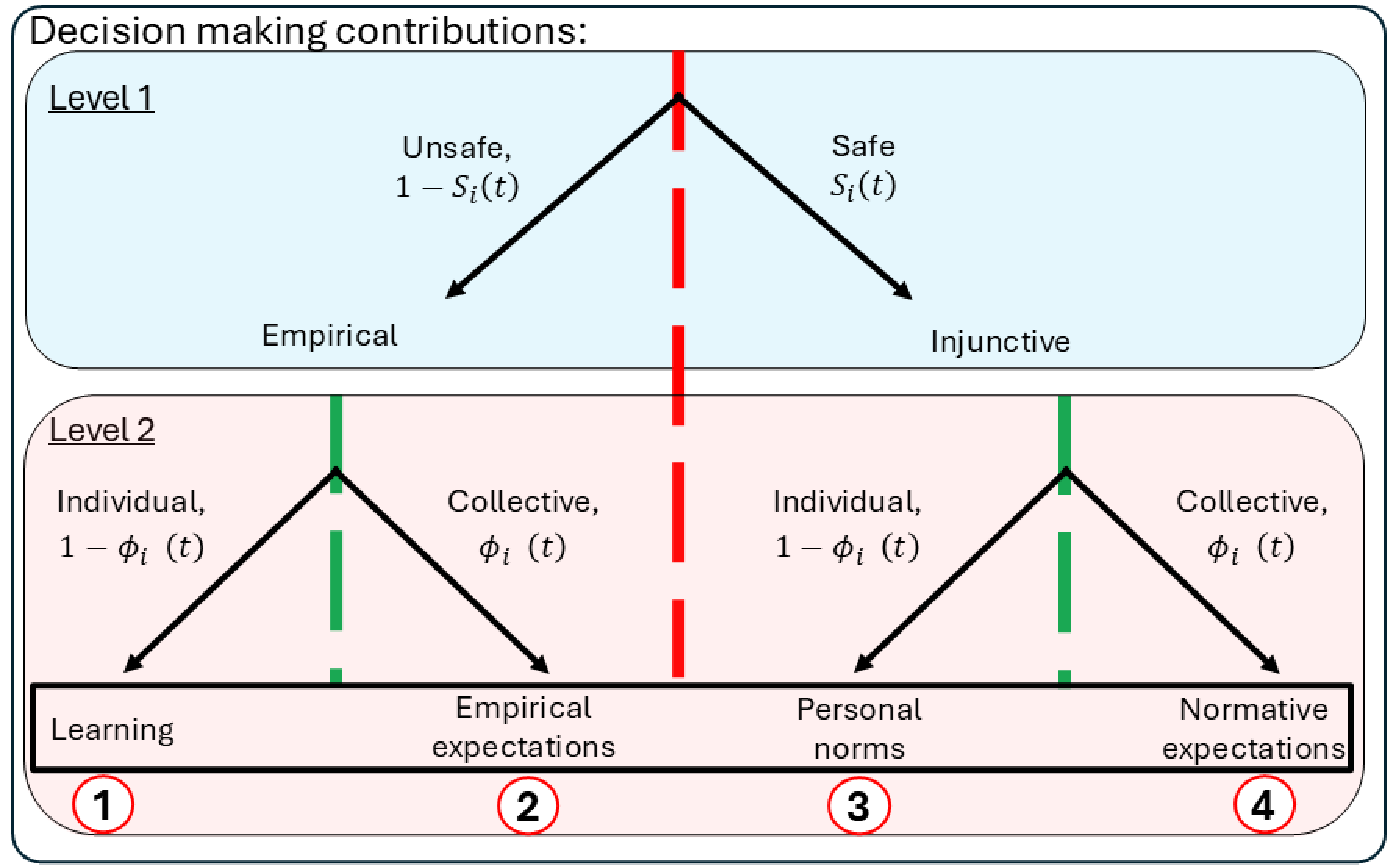}
      \put(0,65){\textbf{a}} % adjust position as needed
    \end{overpic}
    \captionsetup{justification=centering}
    \caption{Decision-making process.}
    \label{fig:Fig2a}
  \end{subfigure}%
  %\hfill
  % ---- Subfigure (b)
  \begin{subfigure}[t]{0.475\textwidth}
    \centering
    \begin{overpic}[width=0.75\linewidth]{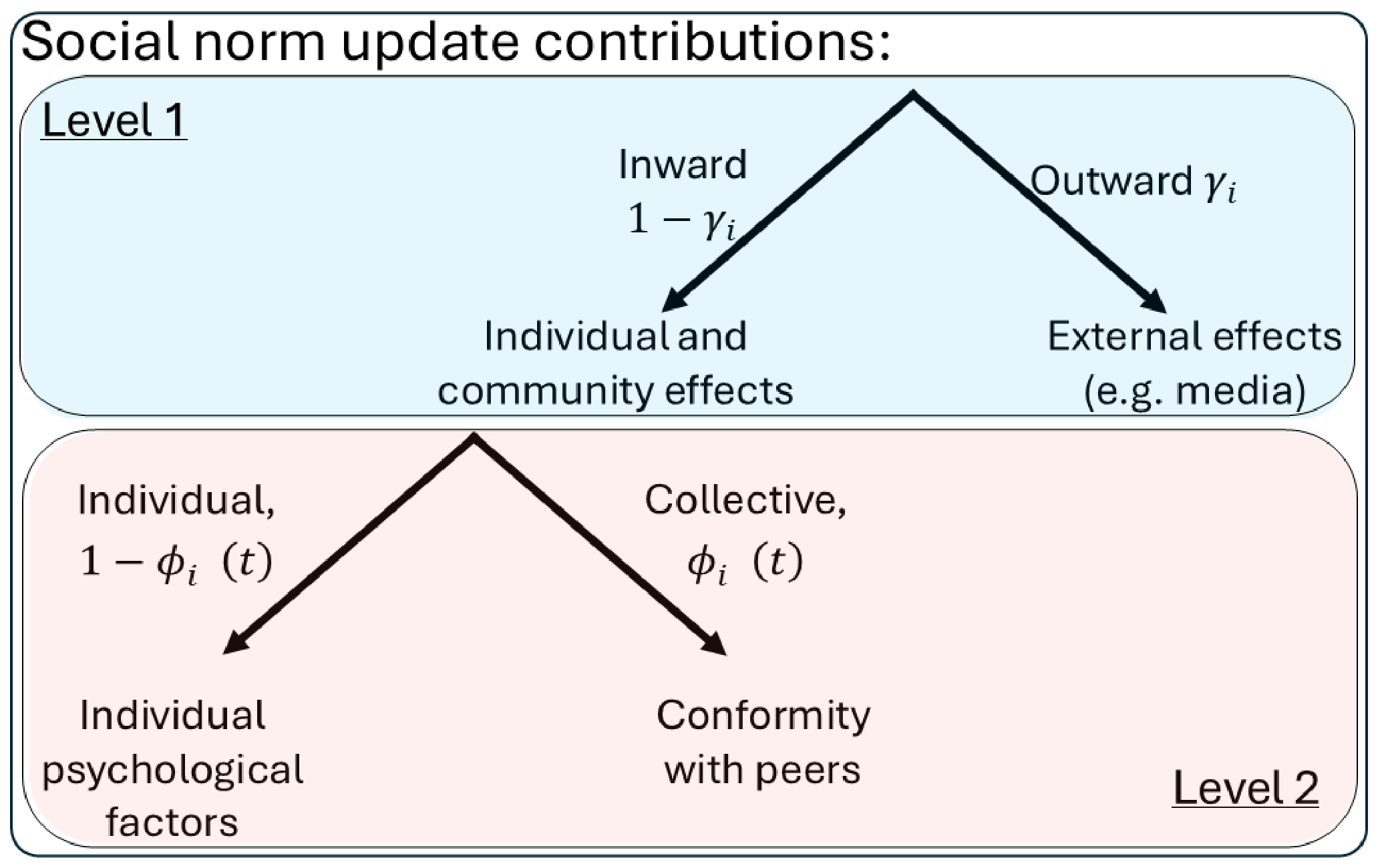}
      \put(0,65){\textbf{b}}
    \end{overpic}
    \captionsetup{justification=centering}
    \caption{Social norms' update.}
    \label{fig:Fig2b}
    \vspace{-10em}
  \end{subfigure}

  \caption{\textbf{Decision-making process and social norm dynamics. a.: Schematic of the decision-making process.} Vaccination intention combines learning (1), personal norms (3), and social norms (2) and (4), each weighted by its relative influence. Agents rely more on empirical factors when they feel unsafe (low $S_i(t)$) or when their environment is unstable and heterogeneous, and more on normative factors under stable conditions. \textbf{b.: Schematic of social-norm dynamics}. Each agent updates norms based on external cues (with probability $\gamma_i$), personal beliefs, and peers’ actions. In unstable and highly divided environments, agents rely more on internal factors than on social conformity.}
  \label{fig:Fig2}
\end{figure}

The contribution from the belief dependent part of the agents, which is
the mechanism activated when the agent is not at risk takes the following
form

\begin{equation}
x_{i}^{injunctive}\left(t\right)=\left(1-\phi_{i}\left(t\right)\right)y_{i}\left(t\right)+\phi_{i}\left(t\right)\widetilde{y}_{i}\left(t\right)
\end{equation}
where again the idea is to split the contributions between the effect
of own beliefs and the beliefs of others using $\phi_i(t)$. We assume $0<\widetilde{y}_{i}\left(t\right),y_{i}\left(t\right)<1$. The dynamics of the norm variables, $y_{i}\left(t\right)$ and $\widetilde{y}_{i}\left(t\right)$ are also defined in the next subsections.
A diagram with the contributions to the total intention is shown in
Fig. \ref{fig:Fig2}.

We employ one of the simplest mechanisms to model whether individuals rely on personal or collective information. We assume that when individuals perceive instability in their neighbors’ behavior and lack of consensus, they rely more on their own judgment, drawing on personal experience or internalized norms. This simplification omits other possible influences, such as the degree of trust in neighbors’ ability to make sound decisions under changing epidemiological conditions, which could depend on factors like disease severity and would substantially complicate the model.

Numerous social and behavioral theories explain how people adapt and change behavior \cite{2015Davis}. Building on the Continuous Opinions and Discrete Actions (CODA) framework \cite{2008Martins}, our approach links individual actions to attitudes shaped by observed behavior, consistent with social-psychological models that emphasize cognitive rules over strict rationality in decision-making \cite{2000Jager}. A similar perspective was used in \cite{2022Tverskoi}. Earlier mathematical models often treated attitudes as static, overlooking their feedback with behavior and thereby misrepresenting adoption dynamics and policy impact.

\setlength{\parskip}{-0.1em}
\subsection*{The experience dependent mechanism\label{subsec:The-experience-dependent}}

Human behaviour combines model-free and model-based reinforcement learning \cite{1999Camerer}, integrating past experience with forward-looking reasoning. The Experience-Weighted Attraction (EWA) model captures both \cite{2013Galla}, unifying reinforcement \cite{2018Sutton} and belief learning \cite{2012Feltovich} by weighting realized and forgone payoffs equally. Forgone payoffs are inferred from neighbours’ average infection rates, since individuals lack direct access to others’ outcomes \cite{2018Sanchez}. Agents evaluate payoffs over the last $m$ cycles with a safety-dependent memory decay. EWA captures how individuals integrate past experience with expectations about others. This empirically validated approach \cite{2018RealpeGomez} provides a cognitively grounded basis for modelling social adaptation and norm compliance.

The payoffs are therefore estimated as following:

\begin{equation}
\begin{array}{c}
\Pi_{i}^{Unvac}\left(t\right)=\begin{cases}
1-c_{I}\widehat{I}_{i,neighbors}\left(t\right) & if\,vaccinaed\\
1-c_{I}\widehat{I}_{i}\left(t\right) & if\,not\,vaccinated
\end{cases}\\
\! \\
\! \\
\Pi_{i}^{Vac}\left(t\right)=1-c_{V}
\end{array}
\end{equation}
and based on this,
the materially-motivated intention of the agent $i$ at time $t$
to vaccinate or not, is assumed to be given by the commonly employed Quantal Response Equilibrium (QRE) for binary choices, which is simply given by the logistic (softmax) function

\begin{equation}
p_{i}^{learn}\left(t\right)=\frac{1}{1+e^{-\frac{\pi_{i}^{Unvac}\left(t\right)-\pi_{i}^{Vac}\left(t\right)}{\kappa}}}\label{eq:learning}
\end{equation}
Here $\pi_{i}^{Vac}\left(t\right)$ is the average payoff for the
vaccinating option over the last $m$ rounds, while $\pi_{i}^{Unvac}\left(t\right)$
is the average payoff received for not vaccinating over the last $m$
rounds $\pi_{i}^{Unvac}\left(t\right)$, given by

\begin{equation}
\begin{array}{c}
\pi_{i}^{Unvac}\left(t\right)=\sum_{j=0}^{m}\frac{\left(S_{i}\left(t\right)\right)^{j}}{\sum_{n=0}^{m}\left(S_{i}\left(t\right)\right)^{n}}\Pi_{i}^{Unvac}\left(t-j\right)\end{array}
\end{equation}
where the safety parameter $S_{i}\left(t\right)$ defined
above plays the role of a memory decaying function. 

\subsection*{Social norm dynamics\label{subsec:Social-norm-dynamics}} To incorporate social norm dynamics into the EWA framework, we separate material payoffs from normative factors. Experiments show that individuals represent others’ preferences and beliefs independently of their own \cite{2002Hedden,2021Szekely}. Accordingly, we model personal attitudes, empirical expectations, and normative expectations as distinct state variables rather than a single imitation process. We are interested in the dynamics of the personal norms $y_i$, normative expectations $\widetilde{y}_i$ and the empirical expectations $\widetilde{x}_i$ which we assume as defined in Gavrilets' model \cite{2021Gavrilets}. 

The dynamics in this work follow a De Groot–type update driven by three psychological forces: cognitive dissonance, social projection, and conformity. Cognitive dissonance aligns actions and self-beliefs \cite{1994Rabin}; social projection reflects the assumption that others are similar to oneself \cite{2007Krueger}; and logical constraints reduce inconsistencies between beliefs about others’ actions and attitudes \cite{2016Friedkin}. Attitudes and expectations thus adapt toward peers’ average behaviour $X_i$ \cite{2015Kashima}, or externally promoted standards $G_i^{l}$. 

Rewriting the equations in \cite{2021Gavrilets}, we obtain
\begin{equation}
\begin{array}{c}
y'_{i}=y_{i}+\xi_{i}^{1}\left[\widehat{C_{i}^{11}}x_{i}+\widehat{C_{i}^{12}}X_{i}+\widehat{C_{i}^{13}}G_{i}^{1}-y_{i}\right]\\
\widetilde{y}'_{i}=\widetilde{y}_{i}+\xi_{i}^{2}\left[\widehat{C_{i}^{21}}y_{i}+\widehat{C_{i}^{22}}X_{i}+\widehat{C_{i}^{23}}G_{i}^{2}-\widetilde{y}_{i}\right]\\
\widetilde{x}'_{i}=\widetilde{x}_{i}+\xi_{i}^{3}\left[\widehat{C_{i}^{31}}\widetilde{y}_{i}+\widehat{C_{i}^{32}}X_{i}+\widehat{C_{i}^{33}}G_{i}^{3}-\widetilde{x}_{i}\right]
\end{array}\label{eq:NormDynamics-1}
\end{equation}
where $\xi_{i}^{l}=\sum_{j}C_{i}^{lj}$ is interpreted as the rate
at which the $l^{th}$ variable ($l=1$ corresponds to personal norms variable $y$, $l=2$ corresponds to normative expectations variable $\widetilde{y}$ and $l=3$ corresponds to empirical expectations variable $\widetilde{x}$) is updated, and $\widehat{C_{i}^{lj}}=\frac{C_{i}^{lj}}{\sum_{k}C_{i}^{lk}}$
such that $\sum_{j}\widehat{C_{i}^{lj}}=1$. In the rest of the article we assume for simplicity that $\xi_{i}^{1}=\xi_{i}^{2}=\xi_{i}^{3}=1$. The results were robust across a wide range of parameter values, in particular we examined the cases where $\xi_{i}^{1}<\xi_{i}^{2}<\xi_{i}^{3}$ which reflects the empirical observations that personal norms are the slowest to adapt while empirical expectations the fastest. This hierarchy is experimentally justified as evidence shows that personal norms remain stable and predictive even when empirical and normative expectations vary, suggesting that personal moral evaluations are less context-dependent than descriptive or injunctive cues \cite{2021Catola}. Future extensions could introduce state-dependent $\xi_i^q$, potentially yielding qualitative transitions similar to those reported in \cite{2024Vriens}.

Expressing the equations in this form clarifies that the coefficients $\widehat{C_{i}^{lj}}$ represent the weights determining how each of the three variables influences the evolution of the $l^{th}$ variable. More specifically, in this formulation, the coefficients $\hat{C}^{\ell 1}_i$ and $\hat{C}^{\ell 2}_i$ can be interpreted as 
self- and peer-weights, respectively: $\hat{C}^{\ell 1}_i$ measures how strongly variable $\ell$ 
(personal norm, normative expectation, or empirical expectation) is pulled towards the agent’s own state, 
whereas $\hat{C}^{\ell 2}_i = 1 - \hat{C}^{\ell 1}_i$ measures the influence of neighbours’ actions. 
When external authorities are present, $\hat{C}^{\ell 3}_i$ captures the weight assigned to externally promoted 
targets $G^{\ell}_i$. In our baseline analysis we set $\hat{C}^{\ell 3}_i = 0$, so that all changes arise 
endogenously from peer influence and internal consistency, and introduce non-zero $\hat{C}^{\ell 3}_i$ 
only when explicitly studying external interventions. Each variable thus depends on two inputs: the individual’s own norms ($x_i$, $y_i$, $\widetilde{y}_i$) and the neighbours’ actions $X_i$. $\widehat{C_{i}^{l1}}$ denotes the weight on self-referential updating, while $1 - \widehat{C_{i}^{l1}}$ captures peer influence, corresponding to inward versus outward psychological orientation. We assume that 
\begin{equation}
C_{i}^{l1}=1-\phi_{i}\left(t\right)
\end{equation}
and hence $C_{i}^{l2}=\phi_{i}\left(t\right)$ where $\phi_{i}\left(t\right)$ as defined in the previous section. This links the coefficients to measurable quantities from the epidemic dynamics rather than assuming arbitrary values, and hence tailors the dynamics to our specific context—the evolution of social norms within a population facing collective epidemic risk. We link these weights to the change-detector $\phi^{change}_{i}\left(t\right)$ and consensus functions $\phi^{consensus}_{i}\left(t\right)$, capturing that agents rely more on personal anchors in unstable environments and on peers in stable ones. This reflects the idea that norm updating accelerates under clear social signals and slows when noise increases (Fig.~\ref{fig:Fig2}B).
\vspace{-1em}
\setlength{\parskip}{-0.1em}

\section{Results}

We run 1000 simulations of the Susceptible-Infected-Recovered
(SIR) model on a network of $N=500$ nodes, using the event-driven algorithm \cite{2017Kiss},
for each season. 

We set the bounded rationality parameter to $\kappa=0.1$, corresponding to a quantal response precision $\lambda=1/\kappa=10$. In quantal response theory (QRT) \cite{2010Goeree}, 
$\lambda$ measures choice sensitivity, higher values imply more deterministic responses, lower values more randomness. This range $\lambda\in[1,10]$ is consistent with laboratory coordination and learning experiments \cite{1999Camerer,2010Goeree}, where individuals choose the higher utility option with $80 - 90\%$ probability. Hence, $\kappa=0.1$ represents a realistic intermediate level of stochasticity, as used in prior behavioral-epidemic models. We adopt $c_V=1$ and $c_I=0.1$ in line with previous game-theoretic and norm-based vaccination models \cite{2004Bauch,2014Oraby,2012dOnofrio,2021Gavrilets}, which typically assume infection to be an order of magnitude more costly than vaccination. Although these parameters are not empirically measured, this ratio reproduces realistic voluntary vaccination levels observed in behavioral experiments \cite{2012Chapman,2022Woike} and reflects the general perception that infection entails far greater health and economic losses than vaccination. Whenever not stated, memory length is assumed to be $m=4$ and $\beta=6$ but results hold qualitatively true for a large range of these parameters as well. Finally for each one of the 1000 simulations a different network realization is assumed, and in all plots lower and upper quartiles are presented. 

\vspace{1em}

\begin{figure}[th!]
\centering

% -------- TOP ROW --------
\begin{overpic}[width=0.32\columnwidth]{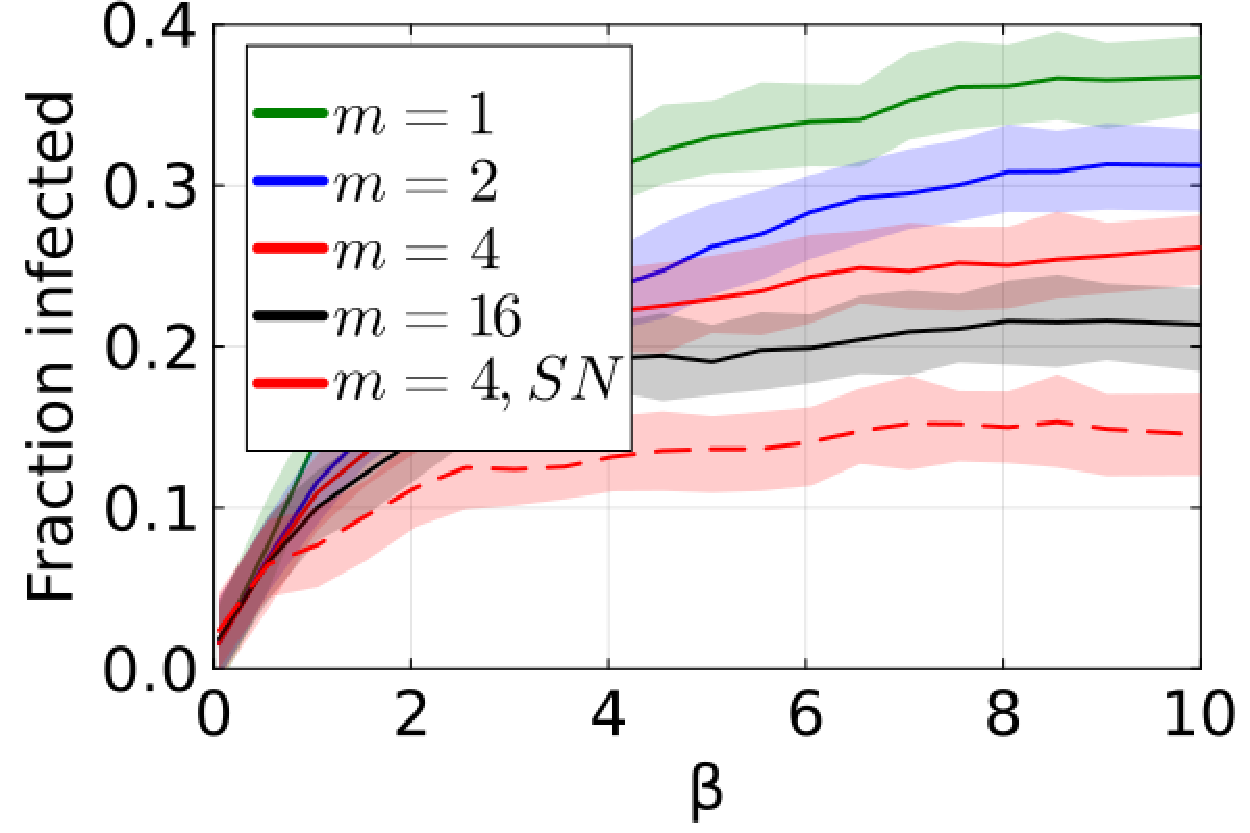}
    \large
    \put(1,69){\textbf{a}}
\end{overpic}
\hfill
\begin{overpic}[width=0.32\columnwidth]{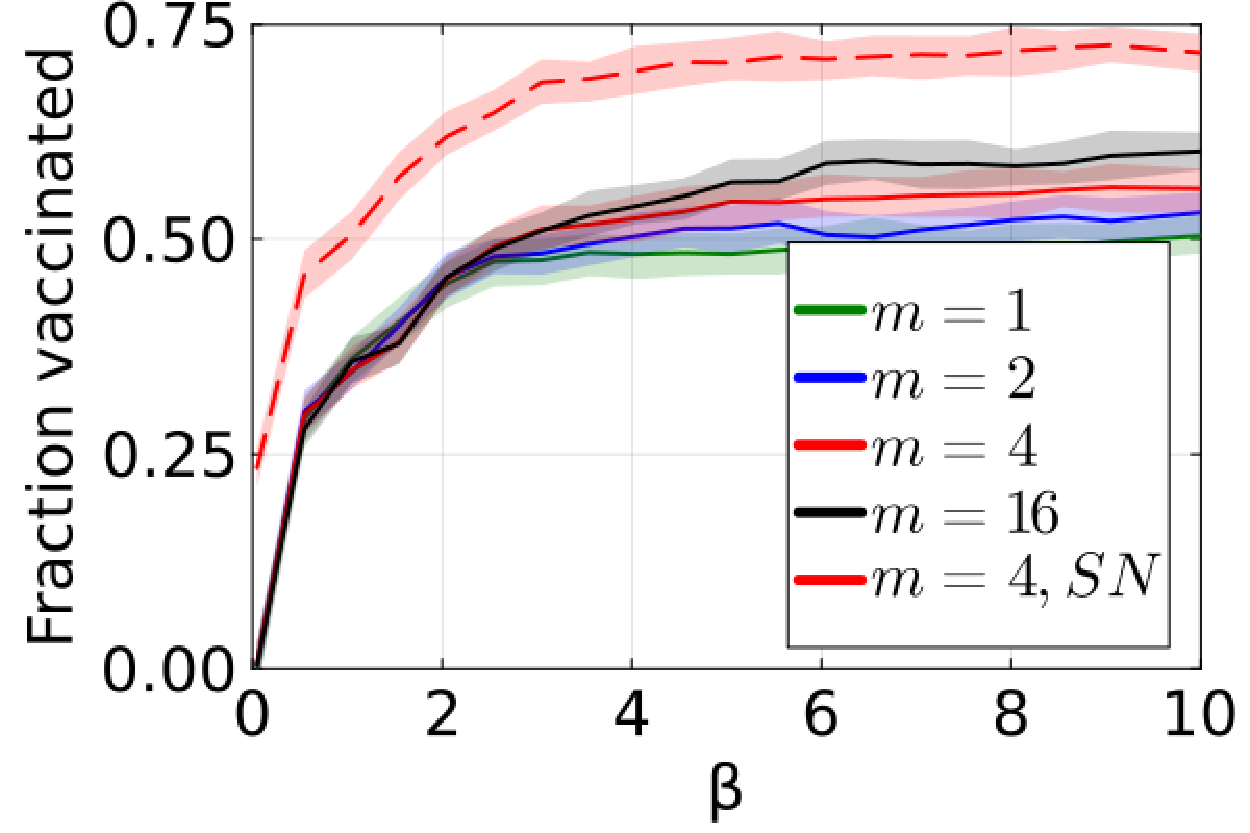}
    \large
    \put(1,69){\textbf{b}}
\end{overpic}
\hfill
\begin{overpic}[width=0.32\columnwidth]{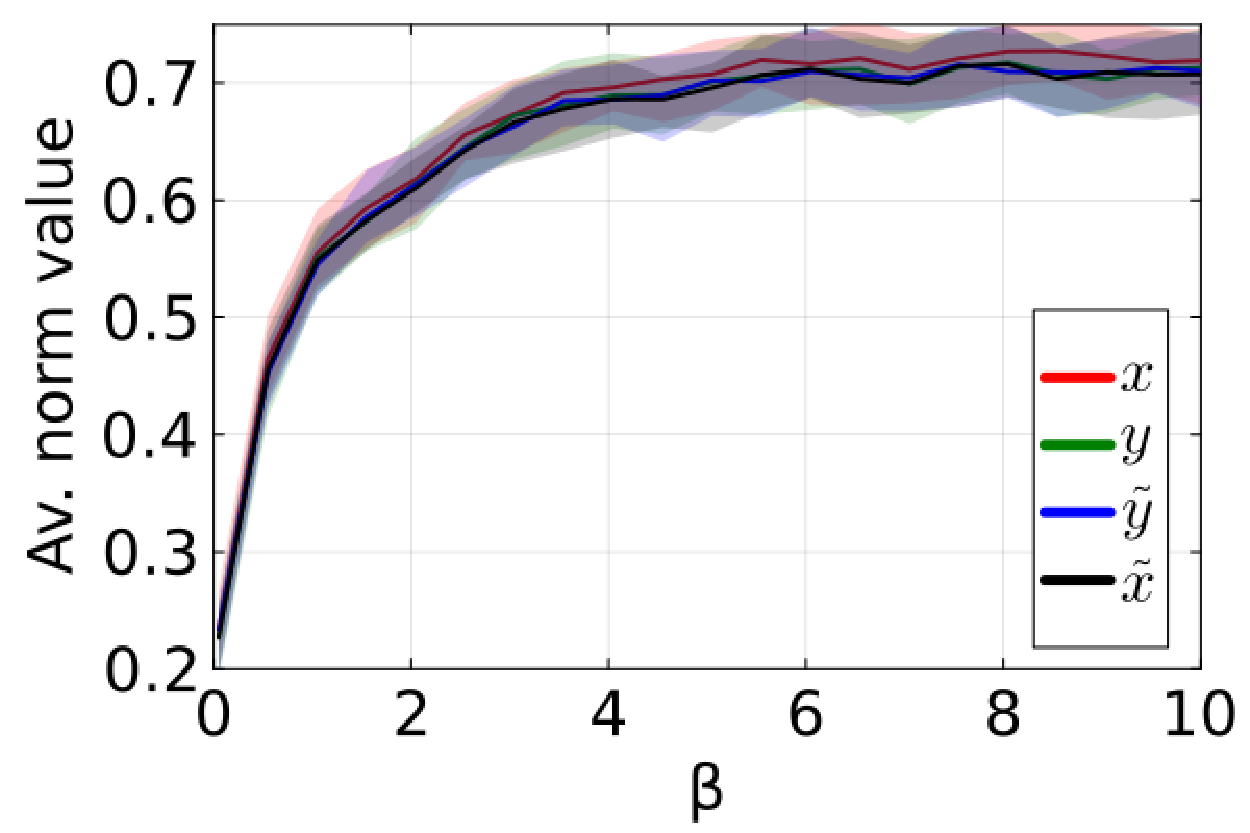}
    \large
    \put(1,69){\textbf{c}}
\end{overpic}

\caption{\label{fig:Fig3} \textbf{Role of social norms in epidemic outcomes.} 
\textbf{a.} Infected fraction versus infectivity rate $\beta$. Infection increases with $\beta$ and saturates at high values, while longer memory reduces outbreak size. For fixed $m=4$, adding social norm dynamics further lowers infections. 
\textbf{b.} Vaccination coverage versus $\beta$. Coverage stabilizes near 0.5 for large $\beta$ due to panic effects, but with norms, even small $\beta$ sustain $\sim20\%$ coverage as heterogeneous initial norms persist. 
\textbf{c.} Mean values of $y$, $\widetilde{y}$, $\widetilde{x}$, and $x$. All converge in the long run, indicating alignment between norms and vaccination intention.}
\end{figure}

The initial values of the three norms associated with each agent are drawn independently from a uniform distribution. Finally, in all the cases studied, we made sure that the system was simulated for sufficiently many seasons such that evolutionary equilibrium was reached, in particular that the percentage of vaccinated individuals did not change by more than 0.025 in the last 50 seasons and with a maximum of 200 seasons. In all of the figures presented in this section we include the lower and upper quartile range as well.
\vspace{-0.0em}
\subsubsection*{The role of social norms\label{subsec:The-role-of}}

Initially, we study the role of memory on epidemic outcomes without considering social norms. Figure~\ref{fig:Fig3}a shows that the infected fraction increases with the infectivity rate $\beta$, remaining close to zero for small $\beta$ and saturating at high values. Longer memory reduces the outbreak size, as agents incorporate a richer history of payoffs when deciding whether to vaccinate. Figure~\ref{fig:Fig3}b displays the corresponding vaccination coverage: for small $\beta$, coverage is close to zero in the absence of norms, while for large $\beta$ it rises toward 0.5, driven by panic effects. Increasing memory boosts coverage further, though not linearly. This is both due to network effects, but also due to the fact that high $\beta$ values reduce the safety factor $S_i(t)$ and thus amplify the role of recent outcomes in shaping decisions. When social norm dynamics are introduced, both infection levels and coverage change significantly. Noteworthy, norms sustain a non-trivial fraction of vaccination (around $20\%$) even at low $\beta$, reflecting the persistence of initial heterogeneous attitudes.
Figure~\ref{fig:Fig3}c illustrates how the three norm variables evolve. As expected, at equilibrium, the three norms coincide, as well as with the vaccination intention.

\subsubsection*{The role of external factors\label{subsec:The-role-of-1}} 

We study the role of an external factor affecting the evolution of
social norms, assuming that the external effect can be present independently
of the epidemiological state of the community. This is introduced
in the following way

\begin{equation}
\begin{array}{c}
y'_{i}=y_{i}+\xi_{i}^{1}\left[\gamma_{i}^{1}G_{i}^{1}+\widetilde{C_{i}^{11}}x_{i}+\widetilde{C_{i}^{12}}X_{i}-y_{i}\right]\\
\widetilde{y}'_{i}=\widetilde{y}_{i}+\xi_{i}^{2}\left[\gamma_{i}^{2}G_{i}^{2}+\widetilde{C_{i}^{21}}y_{i}+\widetilde{C_{i}^{22}}X_{i}-\widetilde{y}_{i}\right]\\
\widetilde{x}'_{i}=\widetilde{x}_{i}+\xi_{i}^{3}\left[\gamma_{i}^{3}G_{i}^{3}+\widetilde{C_{i}^{31}}\widetilde{y}_{i}+\widetilde{C_{i}^{32}}X_{i}-\widetilde{x}_{i}\right]
\end{array}\label{eq:NormDynamicsMedia}
\end{equation}
where $0\leq\gamma_{i}^{l}\leq1$ for $l\in\left\{ 1,2,3\right\} $
is the strength of the external factor signal, and $G_{i}^{l}$ is
the target value for the norms. $\widetilde{C_{i}^{lj}}=\left(1-\gamma_{i}^{l}\right)\widehat{C_{i}^{lj}}$,
where $\widehat{C_{i}^{l1}}$ and $\widehat{C_{i}^{l2}}=1-\widehat{C_{i}^{l1}}$
as before. Note that the condition that the coefficients of the variables
sum up to 1 is still satisfied such that these can be still interpreted
as relative contributions. In our studies, we assume the same strength
$\gamma_{i}^{l}$ and target $G_{i}^{l}$ for all agents, i.e. $\gamma_{i}^{l}=\gamma^{l}$
and $G_{i}^{l}=G^{l}$ $\forall i\in\left\{ 1,...,N\right\} $.

\vspace{0.5em}

\begin{figure}[th!]
\centering

% -------- TOP ROW --------
\begin{overpic}[width=0.48\columnwidth]{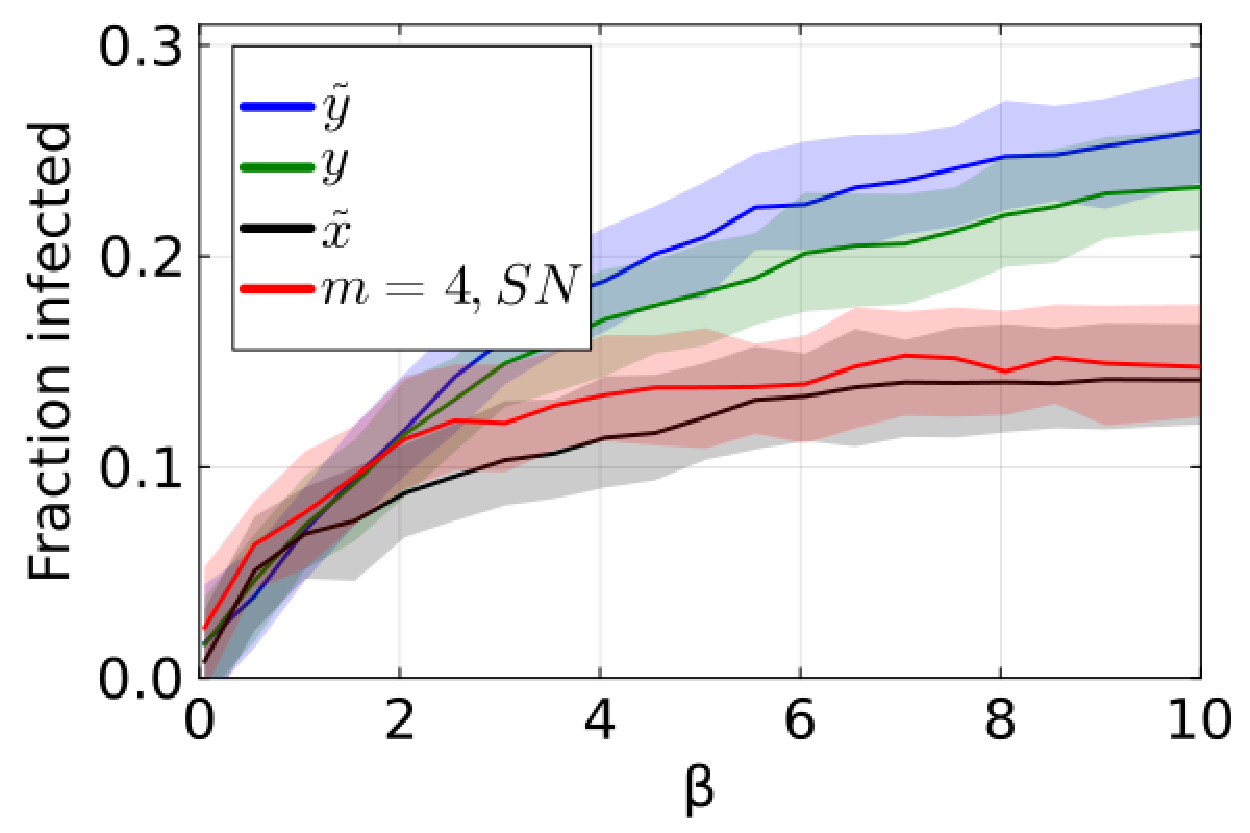}
    \large
    \put(3,68){\textbf{a}}
\end{overpic}
\hfill
\begin{overpic}[width=0.48\columnwidth]{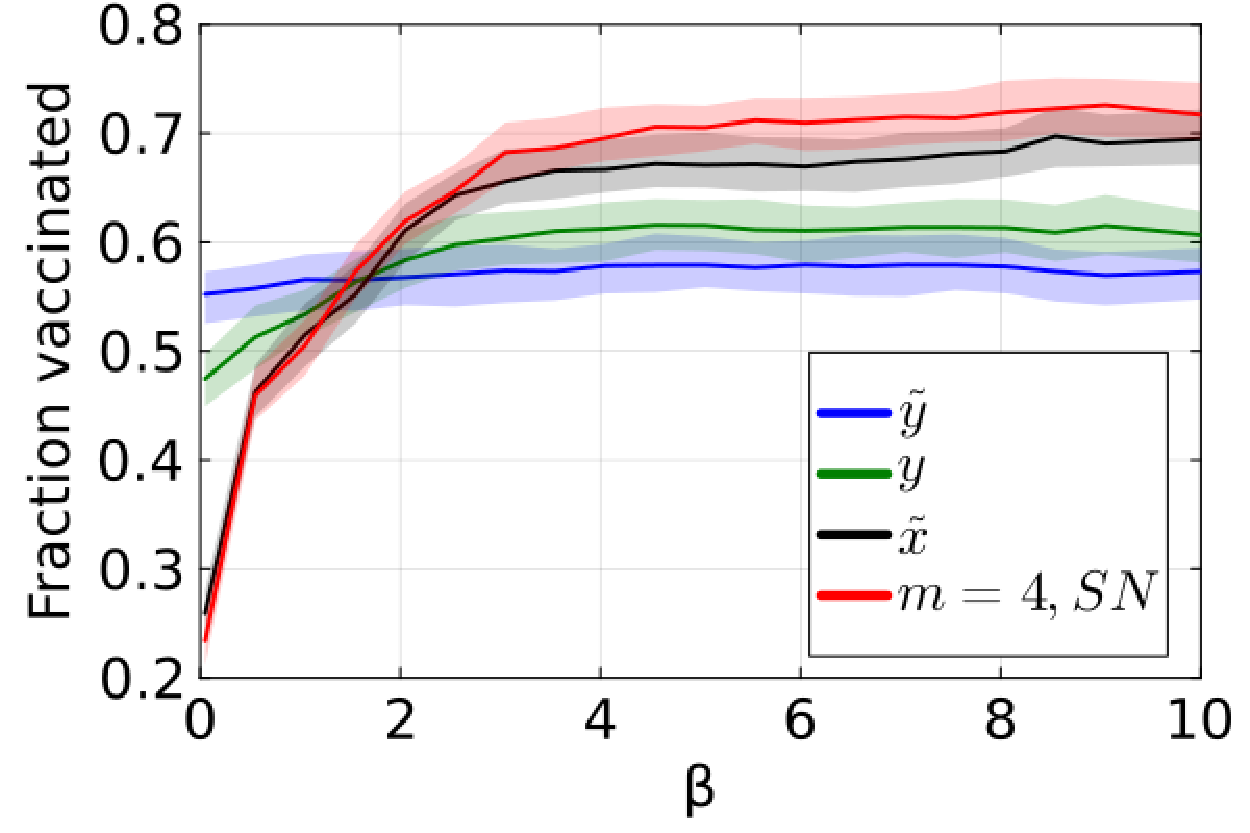}
    \large
    \put(3,68){\textbf{b}}
\end{overpic}
\caption{\label{fig:Fig4} \textbf{Role of external factors on social norms.} 
\textbf{a.} Infected fraction versus infectivity rate $\beta$ when an external factor influences $y_i$, $\widetilde{y}_i$, or $\widetilde{x}_i$. The factor reduces perceived disease severity, driving norms toward $G_i = 0.6$, below the equilibrium value $0.7$ in Fig.~\ref{fig:Fig3}. It increases infections when applied to $y_i$ or $\widetilde{y}_i$, but has negligible effect on $\widetilde{x}_i$. 
\textbf{b.} Vaccination coverage versus $\beta$ for the same cases. Parameters: $\gamma_i = 1.0$, $G_i = 0.6$.}
\end{figure}

\begin{figure}[th!]
\centering

% -------- TOP ROW --------
\begin{overpic}[width=0.48\columnwidth]{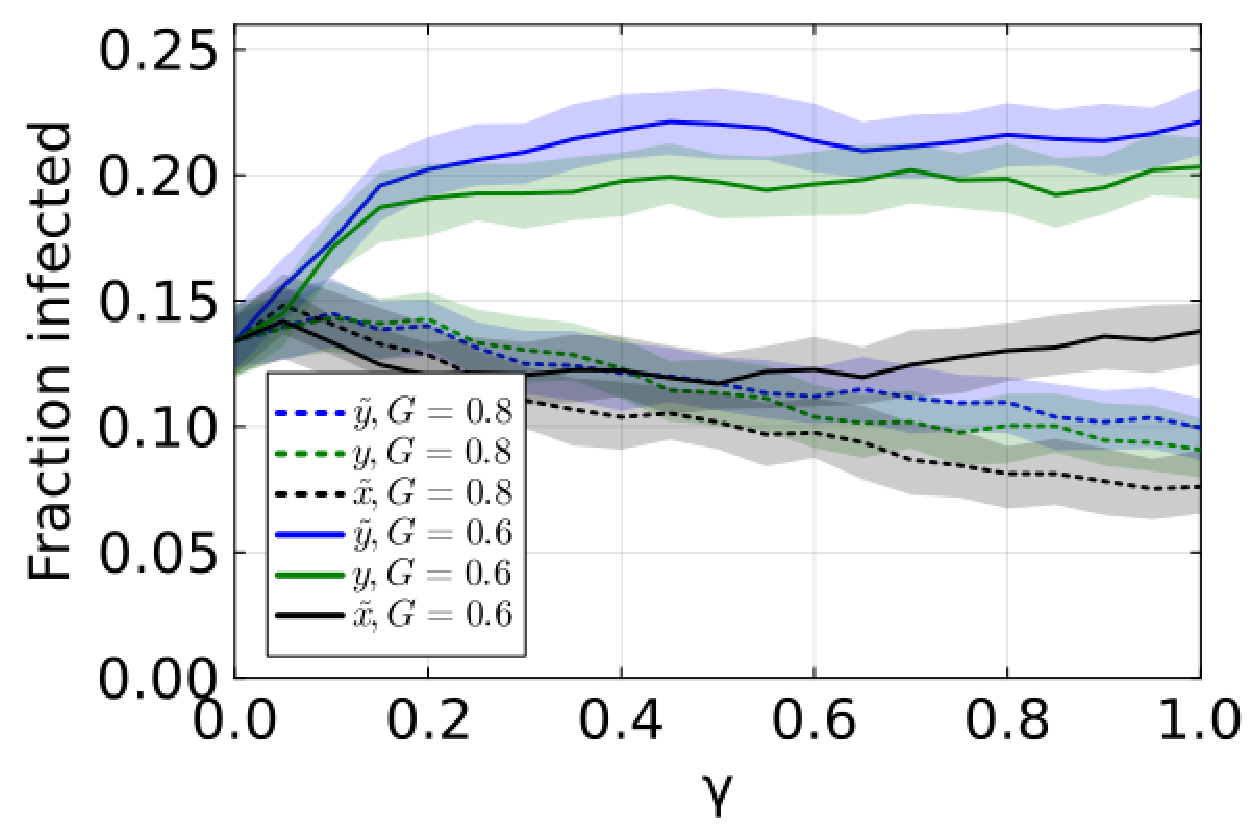}
    \large
    \put(3,66){\textbf{a}}
\end{overpic}
\hfill
\begin{overpic}[width=0.48\columnwidth]{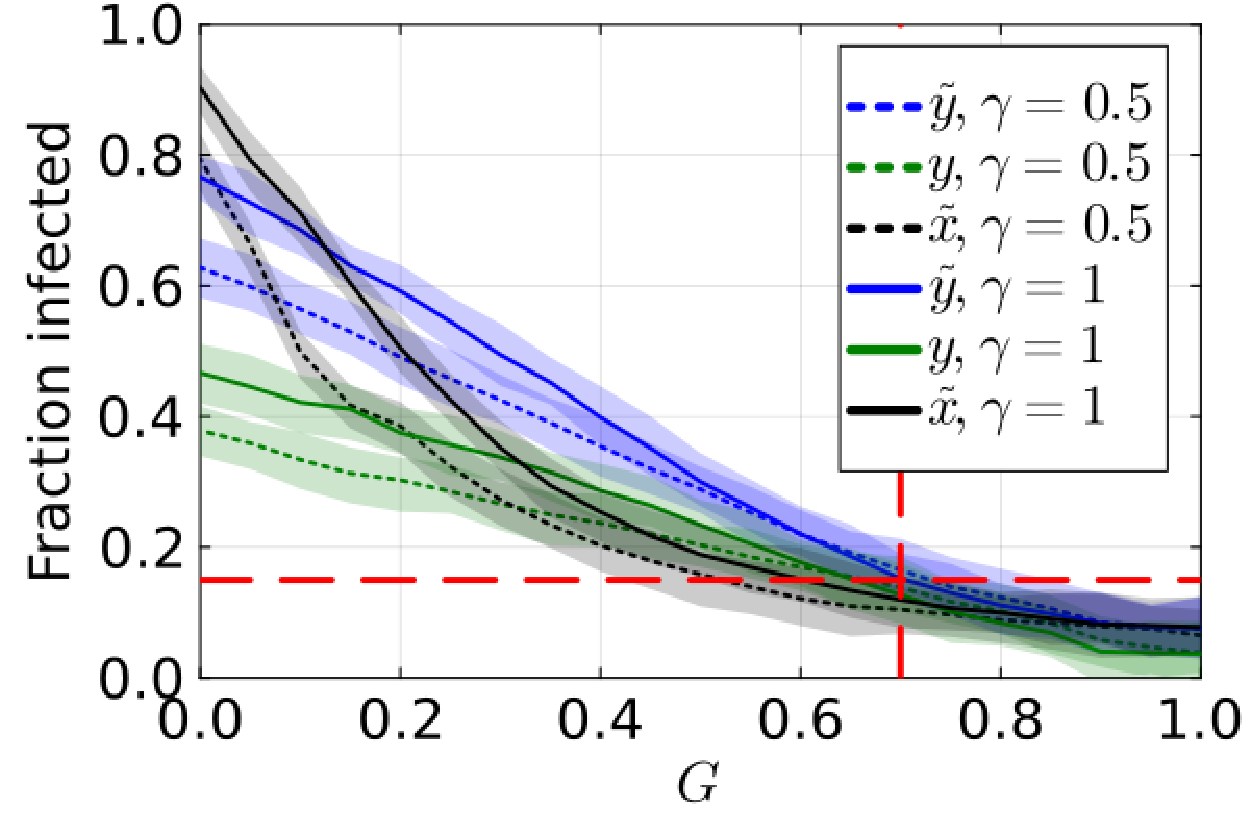}
    \large
    \put(3,66){\textbf{b}}
\end{overpic}
\caption{\label{fig:Fig5} \textbf{Role of the external factor’s strength $\gamma$ and target value $G$.} 
\textbf{a.} Outbreak size versus coupling strength $\gamma$ when the external factor acts on $y_i$, $\widetilde{y}_i$, or $\widetilde{x}_i$, with targets $G = 0.6$ (lower severity perception) and $G = 0.8$ (higher). 
Driving $y_i$ or $\widetilde{y}_i$ toward $G = 0.6$ increases infections, whereas for strong coupling most other cases reduce outbreak size. 
Driving $\widetilde{x}_i$ toward $G = 0.6$ has little effect. 
\textbf{b.} Outbreak size versus $G$ for $\gamma = 0.5$ and $\gamma = 1$. 
The red dashed line marks the baseline without intervention. 
For large $G$ (nudging toward vaccination), infection approaches zero; for small $G$, outbreaks grow. 
Intermediate $G$ values are most effective via $\widetilde{y}$, while for $G \approx 0.6$–$0.8$, interventions on $\widetilde{x}_i$ buffer against external influence.}
\end{figure}

\begin{figure}[th!]
\centering

% -------- TOP ROW --------
\begin{overpic}[width=0.32\columnwidth]{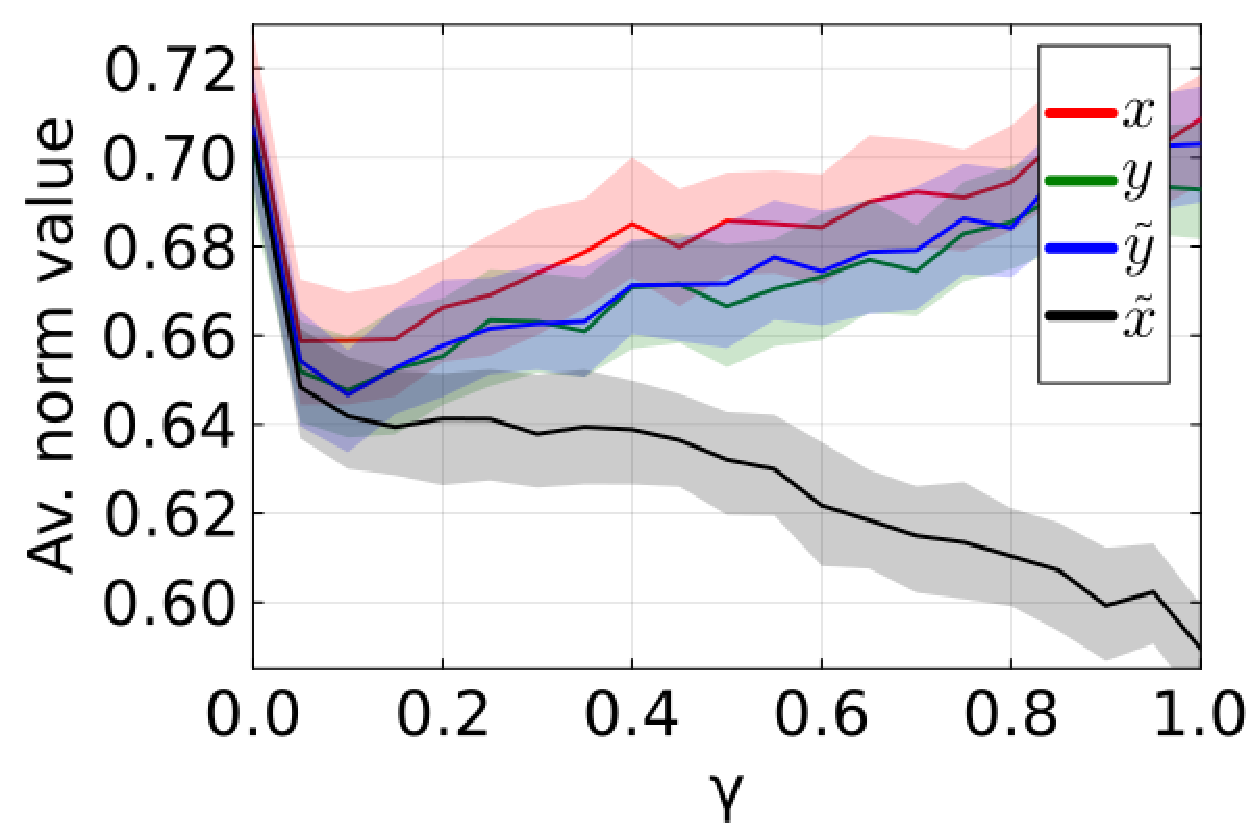}
    \large
    \put(1,69){\textbf{a}}
\end{overpic}
\hfill
\begin{overpic}[width=0.32\columnwidth]{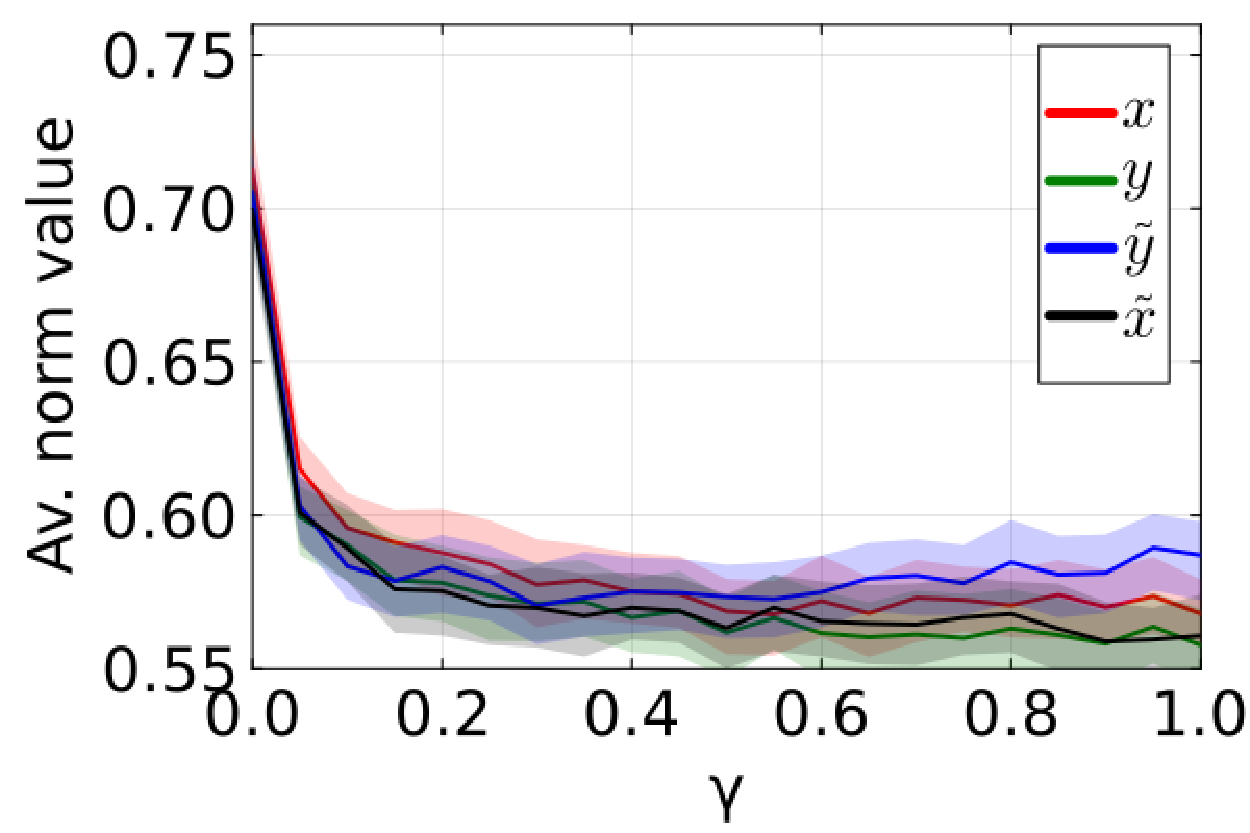}
    \large
    \put(1,69){\textbf{b}}
\end{overpic}
\hfill
\begin{overpic}[width=0.32\columnwidth]{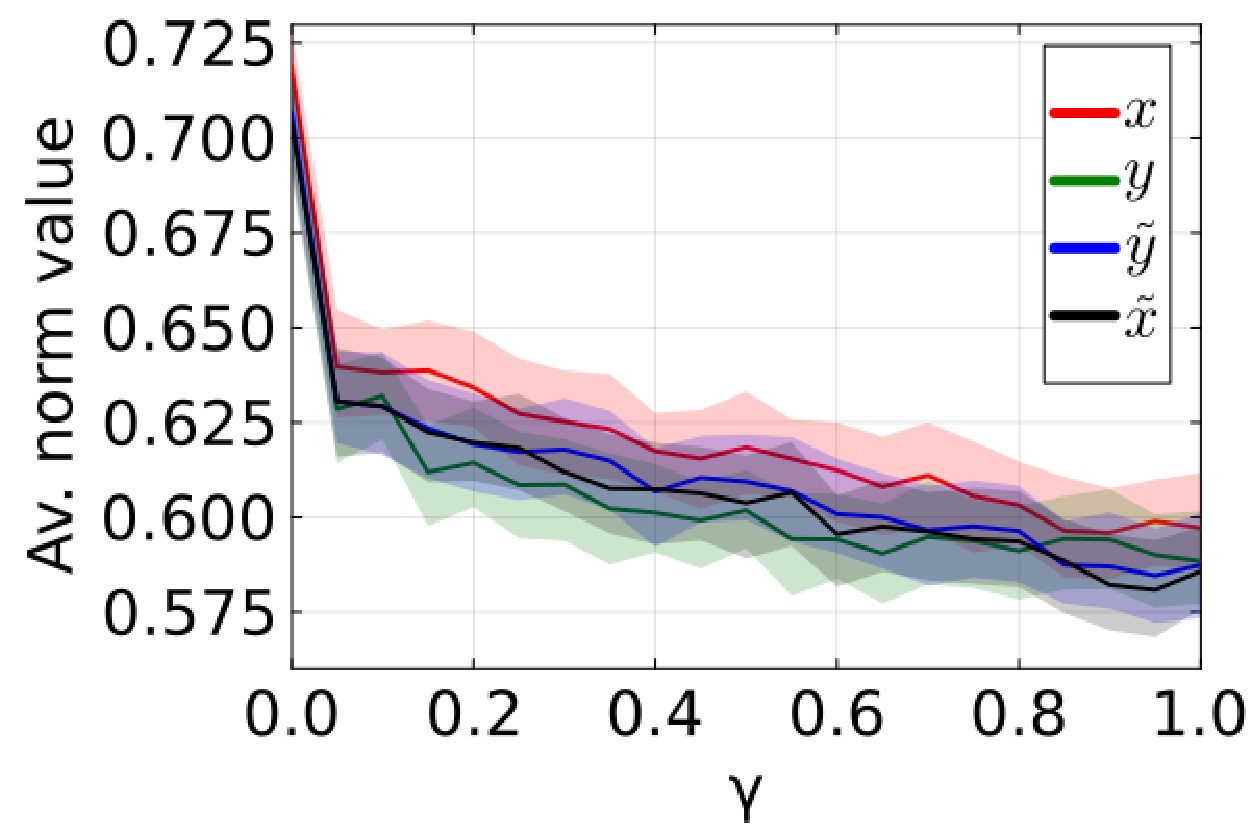}
    \large
    \put(1,69){\textbf{c}}
\end{overpic}
\caption{\label{fig:Fig6} \textbf{Dependence of social norms values on the external factor’s coupling strength $\gamma$.} 
Each scenario applies external coupling to one norm variable ($y$, $\widetilde{y}$, or $\widetilde{x}$) with target value $G = 0.6$. 
Average values across agents are shown. 
\textbf{a.} Coupling to $\widetilde{x}$ drives it toward the target as $\gamma$ increases, while other norms remain unaffected. 
\textbf{b.} Coupling to $\widetilde{y}$ causes all norms to follow its value. 
\textbf{c.} Coupling to $y$ similarly aligns the other norms with it.
}
\end{figure}

Figures ~\ref{fig:Fig4} and ~\ref{fig:Fig5} analyse how external influences on social norms alter epidemic outcomes. Fig.~\ref{fig:Fig4} shows that when an external factor plays down disease severity by shifting norms toward a lower target ($G=0.6$), outbreaks become larger if the intervention acts through personal norms or injunctive expectations, while effects through empirical expectations remain negligible. 
Moving to Fig.~\ref{fig:Fig5}, we observe richer dynamics. Panel~A demonstrates that the outbreak size depends not only on which norm is targeted but also on the strength of the intervention $\gamma$ and its target value $G$. Driving personal norms or injunctive expectations toward lower values reliably increases outbreak size, whereas targeting empirical expectations can sometimes backfire against the external factor: for intermediate values of $G$ ($G\approx 0.6-0.7$), the community resists, and infection levels fall below the baseline without external influence. Panel~B further highlights these nonlinearities, showing that for  target values close to $G \approx 0.7$ outbreaks can shrink under both normative expectation and empirical expectation interventions, with empirical expectations providing the strongest buffer, while for low $G$ values ($G<0.6$) the outbreak size grows as expected. For  $G\approx 0.4-0.6$ targeting personal or injunctive norms remains more effective, however for sufficiently low $G$ targeting empirical expectations might be more effective.  
The nonlinear responses in Fig.~\ref{fig:Fig5} illustrate a boomerang effect \cite{2020Kuang,2016Yakobovitch,2018Richter}: external attempts to weaken vaccination norms sometimes backfire, leading to stronger compliance and reduced infection, akin to backfire effects reported in descriptive norm interventions \cite{2024Rand}.
Finally, Fig.~\ref{fig:Fig6} examines how the average values of each norm evolve under external driving. Only the targeted norm consistently converges to the imposed value, but interventions on injunctive expectations also drag the other norms along, underscoring their coordinating role. 

\subsubsection*{Policy relevance}

Distinguishing injunctive from descriptive norm interventions is essential for effective public-health messaging. Injunctive messages communicate what is socially or morally expected—such as appeals from trusted authorities or statements like “vaccination is a responsible act.” Descriptive messages instead report what most people do (e.g., vaccination rates). Our simulations show that descriptive messages can backfire when they highlight uptake lower than the one without intervention (e.g., $G=0.6$ and $\gamma=1$ for a vaccination uptake without intervention of 0.7), inadvertently signalling that non-vaccination is common. In such cases, individuals conform to majority behavior rather than moral expectations, reducing coverage. Descriptive messaging is thus advisable only when uptake is already high (or paired with injunctive cues emphasizing approval and shared responsibility).

Empirical studies by Heiman et al. \cite{2023Heiman} and Szekely et al. \cite{2021Szekely} confirm that interventions targeting perceived approval and moral norms can strongly shift vaccination intentions, especially under uncertainty about others’ behavior. Our model formalizes how such cues, combined with probabilistic decision-making and social learning, affect collective behavior. While multilayer co-evolutionary frameworks coupling information, vaccination, and epidemics \cite{2022Yin} emphasize structural feedbacks between awareness and contagion, our approach focuses on behavioral feedbacks from evolving normative expectations and adaptive learning. Together, these results show that external influences affect norms asymmetrically: interventions on injunctive norms propagate system-wide, whereas those on empirical expectations can create resilience or even neutralize attempts to downplay risk. This aligns with the experimental evidence of Szekely et al. \cite{2021Szekely} and the theoretical analysis of Roozmand et al. \cite{2022Roozmand}, which identify normative variables as central to norm emergence.

Finally, the simulations highlight three general principles of norm–epidemic coevolution.
(i) Even simple norm dynamics can suppress outbreaks by sustaining vaccination at low infectivity and smoothing behavioral fluctuations, consistent with earlier behavioral–epidemic models where imitation or reinforcement alone cannot maintain cooperation \cite{2004Bauch,2010Reluga}.
(ii) Norm components are asymmetric: external driving applied on normative expectations exerts stronger collective influence than if applied on empirical expectations, consistent with evidence that perceived approval is a key coordination device \cite{2022Woike,2021Szekely} (for not too low values of target values $G$).
(iii) External interventions on norms can have nonlinear and counterintuitive effects \cite{2023Heiman,2023Vriens}. These patterns illustrate how the structure of norm dynamics shapes epidemic and behavioral outcomes, complementing earlier analyses of awareness–disease coupling in multilayer networks \cite{2015Wang,2013Granell}.

\vspace{0.5em}

\section{Discussion}

Epidemic–behavior models often rely on imitation or payoff-based reinforcement, which capture some dynamics but overlook how attitudes, expectations, and social pressures shape preventive choices. To address this, we developed a behavioral epidemic model coupling disease spread with a decision process based on the Experience-Weighted Attraction (EWA) framework \cite{1999Camerer}. EWA integrates reinforcement and belief learning by considering both realized and forgone payoffs, offering a more realistic account of human decision making. We extend this by embedding social norm dynamics, distinguishing between descriptive and injunctive norms \cite{2005Bicchieri,2000Ostrom,2021Gavrilets}. The model captures cognitive dissonance, social projection, and consistency mechanisms that explain how individuals align with peers or authorities.

Our analysis shows that coevolving norm dynamics markedly alter outbreak size. External interventions targeting personal or injunctive expectations strongly affect vaccination coverage, whereas those acting on descriptive norms can be weaker or even counterproductive. Most importantly, the model’s qualitative predictions align with recent experimental findings. 
For example, the strong coordinating role of normative expectations echoes the ``transmission game'' study by Woike \textit{et al.}, 
where injunctive-norm messages reduced risk-taking more effectively than descriptive information or case counts \cite{2022Woike}. 
Likewise, the resilience of norms under external driving and the possibility of boomerang effects parallel evidence that 
normative campaigns can sometimes backfire when they highlight low compliance \cite{2023Heiman,2021Szekely}. 
These parallels suggest a pathway for empirical calibration: longitudinal experiments or panel surveys could track the joint evolution 
of personal norms, empirical expectations, and perceived approval under repeated exposure to norm-based interventions, 
allowing the coefficients in our norm-update equations to be estimated and tested \cite{2023Vriens}. Our model could also be employed to examine the asymmetry in the recovery of social norms observed in Vriens et al. \cite{2024Vriens}.

Our framework is stylized and not predictive, but explanatory, designed to provide theoretical insights on how social norms could potentially coevolve with epidemic risk and shape vaccination behavior. While parameters are not empirically calibrated and agents are homogeneous in risk perception, trust, and norm sensitivity, these simplifications allow a controlled exploration of core behavioral mechanisms. The current setup also assumes fixed social networks and stable information environments, abstracting from media or institutional feedbacks. A recent complementary model incorporating regret and uncertainty \cite{2025Charalambous}, examines the role of more emotional and informational mechanisms in the co-evolving dynamics of disease outbreaks and social norms evolution. Future extensions could incorporate heterogeneous agents, adaptive or multilayer networks \cite{2015Wang,2023Charalambous}, and empirical calibration using experimental or survey data, thereby enhancing realism and policy relevance. Such developments would retain the model’s interpretability while connecting it more closely to real-world interventions. More broadly, empirically grounded norm-based models can guide communication strategies and interventions across collective-action domains, from vaccination and misinformation to environmental cooperation \cite{2017Brewer,2023Moehring,2014Epstein,2022Yin,2006Holme,2017Gavrilets}. Future work could also extend the framework to include homophily in intentions or vaccination status \cite{2006Holme}, and peer sanctions and rewards in norm evolution \cite{2017Gavrilets}.

%\bibliography{Biblio}

\vspace{1em}

\paragraph{Acknowledgements}
I am thankful to Prof. Sandro Meloni for  insightful discussions. This project received funding from the European Union’s Horizon 2020 research and innovation program under the Marie Sklodowska-Curie Grant Agreement No 101034403.

\paragraph{Competing interests} The authors declare no competing financial interests.

\paragraph{Data Availability Statement} This is a computational study and the only data considered is the result of simulations of the coordination game.
To reproduce the research one should recreate the simulations. We facilitate this by describing in detail the
model and tools used in the main text. Replication code can be found in GitHub: \\
\verb+https://github.com/Christos3788/Social-norm-dynamics-in-a-Behavioral-epidemic-model+.

\paragraph{Author Contributions}
Conceptualization: C.C.; Methodology: C.C.; Data curation: C.C. Data visualization: C.C. Writing original draft: C.C.; All authors approved the final submitted draft.

\end{document}